\documentclass[aps,prd,twocolumn,nofootinbib,showpacs]{revtex4}

\usepackage{graphicx,color}
\usepackage[colorlinks=true, linkcolor=blue, citecolor=blue, urlcolor=blue]{hyperref}

\begin{document}

\title{Renormalization Scale Setting for Heavy Quark Pair Production in $e^+e^-$ Annihilation near the Threshold Region}

\author{Sheng-Quan Wang$^{1,2}$}
\email[email:]{sqwang@cqu.edu.cn}

\author{Stanley J. Brodsky$^2$}
\email[email:]{sjbth@slac.stanford.edu}

\author{Xing-Gang Wu$^3$}
\email[email:]{wuxg@cqu.edu.cn}

\author{Leonardo Di Giustino$^{2,4}$}
\email[email:]{leonardo.digiustino@gmail.com}

\author{Jian-Ming Shen$^{5}$}
\email[email:]{cqusjm@cqu.edu.cn}

\address{$^1$Department of Physics, Guizhou Minzu University, Guiyang 550025, P.R. China}
\address{$^2$SLAC National Accelerator Laboratory, Stanford University, Stanford, California 94039, USA}
\address{$^3$Department of Physics, Chongqing University, Chongqing 401331, P.R. China}
\address{$^4$Department of Science and High Technology, University of Insubria, via valleggio 11, I-22100, Como, Italy}
\address{$^5$School of Physics and Electronics, Hunan University, Changsha 410082, P.R. China}

\date{\today}

\begin{abstract}

Heavy fermion pair production in $e^+e^-$ annihilation is a fundamental process in hadron physics and is of considerable interest for various phenomena. In this paper, we will apply the Principle of Maximum Conformality (PMC) to provide a comprehensive analysis of  these processes.
The PMC provides a systematic, unambiguous method for determining the renormalization scales of the QCD  coupling constant for single-scale and multiple-scale applications. The resulting predictions eliminate any renormalization scheme-and-scale ambiguities, eliminate the factorial renormalon divergences, and are consistent with the requirements of the renormalization group.  It is remarkable that two distinctly different scales are determined by using the PMC for heavy fermion pair production near the threshold region. One scale is the order of the fermion mass $m_f$, which enters the hard virtual corrections, and the other scale is of order $ v\,m_f$, where $v$ is the quark velocity, which enters the Coulomb rescattering amplitude. The PMC scales yield the correct physical behavior and reflect the virtuality of the propagating gluons (photons) for the QCD (QED) processes. Moreover, we demonstrate the consistency of PMC scale setting from QCD to QED. Perfect agreement between the Abelian unambiguous Gell-Mann-Low and the PMC scale-setting methods in the limit of zero number of colors is demonstrated.


\end{abstract}

\maketitle

\section{Introduction}
\label{sec:1}

Heavy fermion pair production in $e^+e^-$ annihilation is a fundamental process in the Standard Model (SM). The threshold region is of particular interest. For example, the precise prediction of the production cross section for $e^+e^-\rightarrow\tau^+\tau^-$ in the threshold region is important in order to improve the measurement of the $\tau$-lepton mass~\cite{Asner:2008nq}. Precise theoretical predictions for the production cross section of $e^+e^-\rightarrow c\bar{c}/b\bar{b}$ at the thresholds are  crucial for determining accurate values for  the charm and bottom quark masses, as well as the the QCD coupling constant $\alpha_s$; e.g., as determined from the sum rule method~\cite{Novikov:1976tn, Novikov:1977dq, Voloshin:1995sf}. One of the most important physics goals of future high energy electron-positron colliders is the precise measurement of properties of the top quark,  especially the top quark mass and its width near the threshold region~\cite{Seidel:2013sqa}. A crucial input is the precise prediction of the top quark pair production cross section.

An essential feature of heavy quark pair production in the threshold region of $e^+e^-$ annihilation is the presence of singular terms from the QCD Coulomb corrections. Physically, the renormalization scale which reflects the subprocess virtuality should become very soft in this region. It is conventional to set the renormalization scale to the mass of the heavy fermion $\mu_r=m_f$. This conventional procedure obviously violates the physical behavior of the QCD corrections and will lead inevitably to unreliable predictions for the production cross sections in the threshold region. The resummation of logarithmically enhanced terms is thus required.

It is often argued that one should set the renormalization scale as the typical momentum scale of the process with the purpose of eliminating the large logarithms; this guessed scale is then varied over an arbitrary range to ascertain its uncertainty. However, this conventional procedure gives scheme-dependent predictions, and it thus violates the fundamental principle of renormalization group invariance. The resulting nonconformal perturbative QCD series also has renormalon  n-factorial divergences; one thus introduces inherent renormalization scheme-and-scale uncertainties.    One often argues that the renormalization scale uncertainty by guessing the initial scale will be suppressed by including enough higher-order terms; however, the scale uncertainties become increasingly large at each order, and the renormalon contributions such as $n!\beta^n_0\alpha^n_s$ prevent convergence. One also cannot decide whether poor pQCD convergence is an intrinsic property of the pQCD series, or is simply due to the improper choice of the scale.

In contrast to pQCD, the renormalization scale in Quantum Electrodynamics (QED) is set unambiguously by using the Gell-Mann-Low method~\cite{GellMann:1954fq} where the renormalization scales are set by the virtuality of each photon propagator;  this automatically sums all the proper and improper vacuum polarization contributions  to each photon propagator to all orders. Note that the conventional scale setting method used for pQCD  is incorrect when applied to the Abelian QED theory.  In fact, a correct scale-setting method in pQCD must reduce in the Abelian limit $N_C\rightarrow0$ to the Gell-Mann-Low method~\cite{Brodsky:1997jk}.

The Principle of Maximum Conformality (PMC)~\cite{Brodsky:2011ta, Brodsky:2012rj, Brodsky:2011ig, Mojaza:2012mf, Brodsky:2013vpa} provides a systematic way to eliminate renormalization scheme-and-scale ambiguities. The PMC determines the renormalization scales by absorbing all the $\{\beta_i\}$-terms that govern the behavior of the running coupling via the renormalization group equation. The resulting pQCD series matches the conformal series with $\beta=0$; i.e., it is maximally conformal. Since the PMC predictions do not depend on the choice of the renormalization scheme, PMC scale setting satisfies the principles of renormalization group invariance~\cite{Brodsky:2012ms, Wu:2014iba, Wu:2019mky}. The PMC provides the underlying principle for the well-known Brodsky-Lepage-Mackenzie (BLM) method~\cite{Brodsky:1982gc}, and generalizes the BLM procedure at all orders. By applying PMC scale-setting, the divergent renormalon series disappear, and the convergence of pQCD series is greatly improved.

The PMC approach has been successfully applied to various high energy processes. Recently, we have shown that the correct physical behavior can be obtained using PMC scale setting for the event-shape observables such as the thrust $T$ in electron-positron annihilation~\cite{Wang:2019ljl, Wang:2019isi, DiGiustino:2020fbk}. The PMC scale is not a single fixed value, but it depends continuously on the value of the event-shape observable, reflecting the virtuality of the QCD dynamics.  Thus one can determine the QCD running coupling $\alpha_s(Q^2) $ over a large range of $Q^2$ from a single measurement of $e^+ e^- \to Z^0 \to X$ at $\sqrt{s}=M_Z$.

In this paper, we shall apply the PMC to make comprehensive analyses for the heavy fermion pair production in $e^+e^-$ annihilation near the threshold region. We will show that two distinctly different scales are determined for the heavy fermion pair production near the threshold region. We also will demonstrate the consistency of PMC scale setting in the QED limit.

The remaining sections of this paper are organized as follows. In Sec.\ref{sec:2}, we calculate the QCD process of the quark pair production in $e^+e^-$ annihilation near the threshold region in both the modified minimal subtraction scheme ($\overline{\rm MS}$ scheme) and the V-scheme. In Sec.\ref{sec:3}, we calculate the QED process of the lepton pair production in $e^+e^-$ annihilation near the threshold region. Section \ref{sec:4} is reserved for a summary.

\section{The heavy quark pair production near the threshold region}
\label{sec:2}

\subsection{The QCD process of the quark pair production in the $\overline{\rm MS}$ scheme }
\label{sec:21}

The quark pair production cross section for $e^{+}e^{-}\rightarrow\gamma^{*}\rightarrow Q\bar{Q}$ at the two-loop level can be written as
\begin{eqnarray}
\sigma=\sigma^{(0)}\left[1 + \delta^{(1)}\,a_s(\mu_r) + \delta^{(2)}(\mu_r)\,a^2_s(\mu_r) + {\cal O}(a^3_s)\right],
\label{sigma:1}
\end{eqnarray}
where $a_s(\mu_r)={\alpha_s(\mu_r)}/{\pi}$, $\mu_r$ is the renormalization scale. The LO cross section is
\begin{eqnarray}
\sigma^{(0)}=\frac{4}{3}\frac{\pi\,\alpha^2}{s}N_c\,e^2_Q\frac{v\,(3-v^2)}{2},
\label{qqpairLOpmc}
\end{eqnarray}
and the quark velocity $v$ is
\begin{eqnarray}
v=\sqrt{1-\frac{4\,m_Q^2}{s}}.
\end{eqnarray}
Here, $N_c$ is the number of colors, $e_Q$ is the $Q$ quark electric charge, $s$ is the center-of-mass energy squared and $m_Q$ is the mass of the quark $Q$. The one-loop correction $\delta^{(1)}$ near the threshold region can be written as
\begin{eqnarray}
\delta^{(1)}=C_F\left(\frac{\pi^2}{2\,v}-4\right).
\end{eqnarray}
The two-loop correction $\delta^{(2)}$ can be conveniently split into terms proportional to various $SU(3)$ color factors,
\begin{eqnarray}
\delta^{(2)} &=& C_F^2\,\delta^{(2)}_A + C_F\,C_A\,\delta^{(2)}_{NA} \nonumber\\
&& + C_F\,T_R\,n_f\,\delta^{(2)}_L + C_F\,T_R\,\delta^{(2)}_H.
\end{eqnarray}
The terms $\delta^{(2)}_{A}$, $\delta^{(2)}_{L}$ and $\delta^{(2)}_{H}$ are the same in either Abelian or non-Abelian theories; the term $\delta^{(2)}_{NA}$ only arises in the non-Abelian theory. This process provides the opportunity to explore rigorously the scale-setting method in the non-Abelian and Abelian theories.

The Coulomb correction plays an important role in the threshold region; it is proportional to powers of $(\pi/v)$. The renormalization scale is thus  relatively soft in this region. In fact,  the PMC scales must be determined separately for the non-Coulomb and Coulomb corrections~\cite{Brodsky:1995ds, Brodsky:2012rj}. When the quark velocity $v\rightarrow0$, the Coulomb correction dominates the contribution of the production cross section, and the contribution of the non-Coulomb correction will be suppressed. On general grounds one expects that threshold physics is governed by the nonrelativistic Coulomb instantaneous potential. The potential affects the cross section through final state interactions when the scale is above threshold; it leads to bound states when the scale is below threshold.

The cross section given in Eq.(\ref{sigma:1}) is further divided into the $n_f$-dependent and $n_f$-independent parts, i.e.,
\begin{widetext}
\begin{eqnarray}
\sigma &=&\sigma^{(0)}\left[1+\delta^{(1)}_h\,a_s(\mu_r) + \left(\delta^{(2)}_{h,in}(\mu_r)+\delta^{(2)}_{h,n_f}(\mu_r)\,n_f\right)\,a^2_s(\mu_r) \right. \nonumber\\
&&\left. + \left(\frac{\pi}{v}\right)\,\delta^{(1)}_{v}\,a_s(\mu_r) + \left(\frac{\pi}{v}\right)\,\left(\delta^{(2)}_{v,in}(\mu_r)+\delta^{(2)}_{v,n_f}(\mu_r)\,n_f\right)\,a^2_s(\mu_r)+ \left(\frac{\pi}{v}\right)^2\,\delta^{(2)}_{v^2}\,a^2_s(\mu_r) + {\cal O}(a^3_s)\right].
\label{eqpairNNLO}
\end{eqnarray}
\end{widetext}
The coefficients $\delta^{(1)}_h$ and $\delta^{(2)}_h$ are for the non-Coulomb corrections, and the coefficients $\delta^{(1)}_v$, $\delta^{(2)}_v$ and $\delta^{(2)}_{v^2}$ are for the Coulomb corrections. These coefficients in the $\overline{\rm MS}$ scheme are calculated in Refs.~\cite{Czarnecki:1997vz, Beneke:1997jm, Bernreuther:2006vp} and at the scale $\mu_r=m_Q$ they can be written as
\begin{eqnarray}
\delta^{(1)}_h=-4\,C_F,~\delta^{(1)}_{v}=\frac{C_F\,\pi}{2},
\end{eqnarray}
\begin{eqnarray}
\delta^{(2)}_{h,in}&=&-{1\over72}\,C_F\,(C_A\,(302+468\zeta_3 + \pi^2(-179+192\ln2)) \nonumber\\
&& - 2\,(-16(-11+\pi^2)\,T_R + C_F\,(351+6\pi^4-36\zeta_3 \nonumber\\
&& + \pi^2\,(-70+48\ln2))) + 24(3\,C_A+2\,C_F)\,\pi^2\ln v), \nonumber\\
\delta^{(2)}_{h,n_f}&=&{11\,C_F\,T_R\over9}, \nonumber\\
\delta^{(2)}_{v,in}&=&-{1\over72}\,C_F\,\pi\,(-31\,C_A+144\,C_F+66\,C_A\ln(2v)), \nonumber\\
\delta^{(2)}_{v,n_f}&=&{1\over18}\,C_F\,\pi\,T_R\,(-5+6\ln(2v)), \nonumber\\
\delta^{(2)}_{v^2}&=&{C_F^2\,\pi^2\over12}.
\end{eqnarray}
After absorbing the nonconformal term $\beta_0=11/3\,C_A-4/3\,T_R\,n_f$ into the coupling constant using the PMC, we obtain
\begin{eqnarray}
\sigma &=& \sigma^{(0)}\left[1+\delta^{(1)}_h\,a_s(Q_h) + \delta^{(2)}_{h,\rm sc}(\mu_r)\,a^2_s(Q_h) \right. \nonumber\\
&&\left. + \left(\frac{\pi}{v}\right)\,\delta^{(1)}_v\,a_s(Q_v) + \left(\frac{\pi}{v}\right)\,\delta^{(2)}_{v,\rm sc}(\mu_r)\,a^2_s(Q_v)\right. \nonumber\\
&&\left. + \left(\frac{\pi}{v}\right)^2\,\delta^{(2)}_{v^2}\,a^2_s(Q_v) + {\cal O}(a^3_s)\right].
\label{eqpairNNLOpmc}
\end{eqnarray}
The PMC scales $Q_i$ can be written as
\begin{eqnarray}
Q_i = \mu_r\exp\left[\frac{3\,\delta^{(2)}_{i,n_f}(\mu_r)}{2\,T_R\,\delta^{(1)}_i}\right],
\end{eqnarray}
and the coefficients $\delta^{(2)}_{i,\rm sc}(\mu_r)$ are
\begin{eqnarray}
\delta^{(2)}_{i,\rm sc}(\mu_r) = \frac{11\,C_A\,\delta^{(2)}_{i,n_f}(\mu_r)}{4\,T_R}+\delta^{(2)}_{i,in}(\mu_r),
\end{eqnarray}
where, $i=h$ and $v$ stand for the non-Coulomb and Coulomb corrections, respectively. The nonconformal $\beta_0$ term is eliminated, and the resulting pQCD series matches the conformal series and thus only the conformal coefficients remain in the cross section. The conformal coefficients are independent of the renormalization scale $\mu_r$. At the present two-loop level, the PMC scales are also independent of the renormalization scale $\mu_r$. Thus, the resulting cross section in Eq.(\ref{eqpairNNLOpmc}) eliminates the renormalization scale uncertainty.

Taking $C_A=3$, $C_F=4/3$ and $T_R=1/2$ for QCD, the PMC scales in the $\overline{\rm MS}$ scheme are
\begin{eqnarray}
Q_h =e^{(-11/24)}\,m_Q
\label{scaleQCDhms}
\end{eqnarray}
for the non-Coulomb correction, and
\begin{eqnarray}
Q_v =2\,e^{(-5/6)}\,v\,m_Q
\label{scaleQCDbms}
\end{eqnarray}
for the Coulomb correction. The scale $Q_h$ originates from the hard gluon virtual corrections, and thus it is determined for the short-distance process. The scale $Q_v$ originates from Coulomb rescattering. Since the PMC scales are determined by absorbing the nonconformal $\{\beta_i\}$-terms, the behavior of the scale is controlled by the coefficient of the QCD $\beta$ function. It is noted that the coefficient of the $\beta_0$ function for the non-Coulomb correction is independent of the quark velocity $v$, whereas the logarithmic term $\ln(2v)$ appears in the coefficient of the $\beta_0$ function for the Coulomb correction. As expected, the resulting scale $Q_h$ is of the order $m_Q$, whereas the scale $Q_v$ is of the order $v\,m_Q$.

\begin{figure}[htb]
\centering
\includegraphics[width=0.40\textwidth]{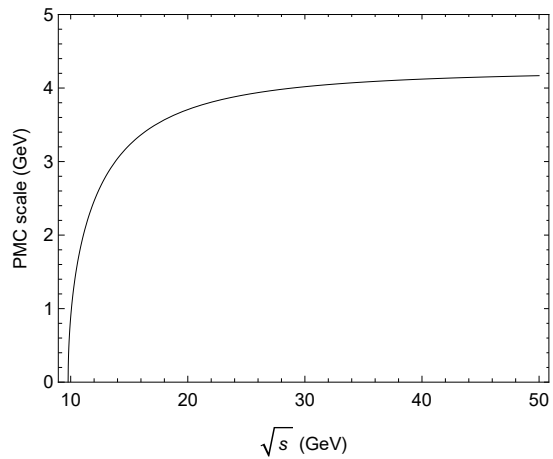}
\caption{The PMC scale $Q_v$ versus the center-of-mass energy $\sqrt{s}$ for the $b$ quark pair production in the $\overline{\rm MS}$ scheme. $m_Q=4.89$ GeV. }
\label{figPMCscalebeta0}
\end{figure}

In the following, we will take the bottom quark pair production as an example to make a detailed analysis near the threshold region. Taking $m_Q=4.89$ GeV~\cite{Bernreuther:2016ccf}, we obtain
\begin{eqnarray}
Q_h = 3.09~\rm GeV,
\end{eqnarray}
which is smaller than $m_Q$ in the $\overline{\rm MS}$ scheme. For the Coulomb correction part, the scale $Q_v$ is shown in Fig.(\ref{figPMCscalebeta0}). It shows that the scale $Q_v$ depends continuously on the quark velocity $v$, and  it becomes soft for $v\rightarrow0$, yielding the correct physical behavior of the scale and reflecting the virtuality of the QCD dynamics. Also the number of active flavors $n_f$ changes with the quark velocity $v$ according to the PMC scale.

When the quark velocity $v\rightarrow0$, the small scale in the coupling constant demonstrates that the perturbative QCD theory becomes unreliable and non-perturbative effects must be taken into account. One can adopt the Light Front Holographic QCD (LFHQCD)~\cite{Brodsky:2014yha} to evaluate the coupling constant $\alpha_s(Q)$ in the low scale region. According to the LFHQCD, the coupling constant $\alpha_s(Q)$ is finite for $Q\rightarrow0$.

In contrast, the renormalization scale is simply fixed at $\mu_r=m_Q$ using conventional scale setting. Our calculations show that in the $\overline{\rm MS}$ scheme, the scale should be $e^{(-11/24)}\,m_Q$, which is smaller than $m_Q$ for the non-Coulomb correction. For the Coulomb correction, since the scale becomes soft for $v\rightarrow0$, simply fixing the renormalization scale $\mu_r=m_Q$ obviously violates the physical behavior and lead to unreliable predictions in the threshold region. The resummation of logarithmically enhanced terms is thus required.

\begin{figure}[htb]
\centering
\includegraphics[width=0.40\textwidth]{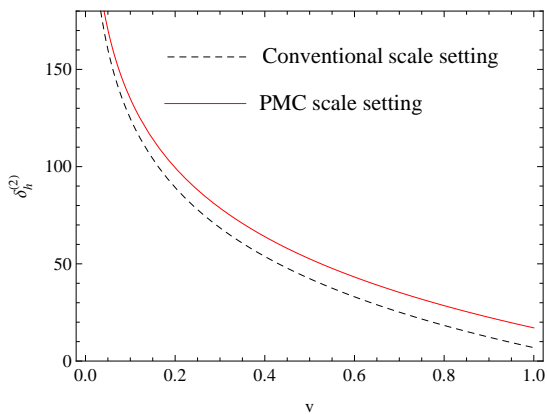}
\caption{The two-loop coefficients $\delta^{(2)}_h$ of the non-Coulomb correction in the $\overline{\rm MS}$ scheme for the $b$ quark pair production, where $\delta^{(2)}_h=(\delta^{(2)}_{h,in}+\delta^{(2)}_{h,n_f}\,n_f)$ is for conventional scale setting while $\delta^{(2)}_h=\delta^{(2)}_{h,\rm sc}$ is for PMC scale setting.}
\label{figCoedetah2}
\end{figure}

We present the two-loop coefficients $\delta^{(2)}_h$ of the non-Coulomb correction in the $\overline{\rm MS}$ scheme using conventional and PMC scale settings in Fig.(\ref{figCoedetah2}). Figure (\ref{figCoedetah2}) show that the $v$-dependent behavior of the coefficients $\delta^{(2)}_h$ is the same but their magnitudes are different using conventional and PMC scale settings. When the quark velocity $v\rightarrow0$, the behavior of the non-Coulomb correction coefficients using conventional and PMC scale settings is divergent $\delta^{(2)}_h\rightarrow+\infty$ due to the presence of the term $-\ln v$. As expected, after multiplying this term by the $v$-factor in the LO cross section $\sigma^{(0)}$ given in Eq.(\ref{qqpairLOpmc}), the contribution of the non-Coulomb corrections is finite and is suppressed near the threshold region, i.e., $(\sigma^{(0)}\,\delta^{(2)}_h\,a^2_s)\rightarrow0$ for the quark velocity $v\rightarrow0$.

\begin{figure}[htb]
\centering
\includegraphics[width=0.40\textwidth]{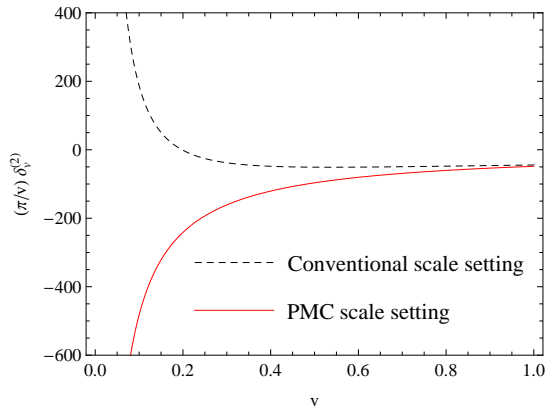}
\caption{The Coulomb terms of the form $(\pi/v)\,\delta^{(2)}_{v}$ in the $\overline{\rm MS}$ scheme for the $b$ quark pair production, where $\delta^{(2)}_v=(\delta^{(2)}_{v,in}+\delta^{(2)}_{v,n_f}\,n_f)$ is for conventional scale setting and $\delta^{(2)}_v=\delta^{(2)}_{v,\rm sc}$ is for PMC scale setting. }
\label{figCoedetab2}
\end{figure}

For the Coulomb correction, the resummation of the Coulomb term of the form $(\pi/v)^2\,\delta^{(2)}_{v^2}$ results in the well known Sommerfeld rescattering formula~\cite{Czarnecki:1997vz}. For the Coulomb term of the form $(\pi/v)\,\delta^{(2)}_{v}$, we present its $v$-dependent behavior in the $\overline{\rm MS}$ scheme using conventional and PMC scale settings in Fig.(\ref{figCoedetab2}). It shows that when the quark velocity $v\rightarrow0$, the $v$-dependent behavior of the Coulomb term $(\pi/v)\,\delta^{(2)}_{v}$ is dramatically different using conventional and PMC scale settings. In the case of conventional scale setting, its behavior is $(\pi/v)\delta^{(2)}_v\rightarrow+\infty$ for $v\rightarrow0$ due to the presence of the term $-\ln v/v$. After multiplying this term by the $v$-factor in the LO cross section $\sigma^{(0)}$, the contribution from the Coulomb term $(\pi/v)\,\delta^{(2)}_v$ using conventional scale setting is not finite, i.e., $(\sigma^{(0)}\,(\pi/v)\,\delta^{(2)}_v)\rightarrow+\infty$ for $v\rightarrow0$. It should be stressed that the term $\ln v$ vanishes, and the term $-1/v$ remains in the conformal coefficient after applying PMC scale setting. Thus the $v$-dependent behavior is $(\pi/v)\delta^{(2)}_v\rightarrow-\infty$ for $v\rightarrow0$. This term $-1/v$ is canceled by multiplying it by the $v$-factor in the LO cross section $\sigma^{(0)}$, and thus the contribution from the Coulomb term $(\pi/v)\,\delta^{(2)}_{v}$ using PMC scale setting is finite for $v\rightarrow0$. It is noted that the contributions of the Coulomb correction using conventional and PMC scale settings are suppressed for $v\rightarrow1$.

\subsection{The QCD process of the quark pair production in the V-scheme }
\label{sec:22}

The quark pair production cross section in the above analysis is calculated in the $\overline{\rm MS}$ scheme. Effective charge $a^V_s=\alpha_V/\pi$ (V-scheme) defined by the interaction potential between two heavy quarks~\cite{Appelquist:1977tw, Fischler:1977yf, Peter:1996ig, Schroder:1998vy, Smirnov:2008pn, Smirnov:2009fh, Anzai:2009tm},
\begin{eqnarray}
V(Q^2) = -{4\,\pi^2\,C_F\,a^V_s(Q)\over Q^2},
\end{eqnarray}
provides a physically-based alternative to the usual $\overline{\rm MS}$ scheme. As in the case of QED, when the scale of the coupling $a^V_s$ is identified with the exchanged momentum, all vacuum polarization corrections are resummed into $a^V_s$. By using the relation between $a_s$ and $a^V_s$ at the one-loop level, i.e.,
\begin{eqnarray}
a^V_s(Q) = a_s(Q) + \left({31\over36}C_A-{5\over9}T_R\,n_f\right)a^2_s(Q) + {\cal O}(a^3_s),
\label{Vscheme:as}
\end{eqnarray}
we convert the quark pair production cross section from the $\overline{\rm MS}$ scheme to the V-scheme. The corresponding perturbative coefficients in Eq.(\ref{eqpairNNLO}) in the V-scheme are
\begin{eqnarray}
\sigma^{(0)}|_V=\sigma^{(0)},
\end{eqnarray}
\begin{eqnarray}
\delta^{(1)}_h|_V=\delta^{(1)}_h,~\delta^{(1)}_{v}|_V=\delta^{(1)}_{v},
\end{eqnarray}
\begin{eqnarray}
\delta^{(2)}_{h,in}|_V&=&\delta^{(2)}_{h,in} - {31\over36}\,C_A\,\delta^{(1)}_h, \nonumber\\
\delta^{(2)}_{h,n_f}|_V&=&\delta^{(2)}_{h,n_f} + {5\over9}\,T_R\,\delta^{(1)}_h, \nonumber\\
\delta^{(2)}_{v,in}|_V&=&\delta^{(2)}_{v,in} - {31\over36}\,C_A\,\delta^{(1)}_{v}, \nonumber\\
\delta^{(2)}_{v,n_f}|_V&=&\delta^{(2)}_{v,n_f} + {5\over9}\,T_R\,\delta^{(1)}_{v}, \nonumber\\
\delta^{(2)}_{v^2}|_V&=&\delta^{(2)}_{v^2}.
\end{eqnarray}
After applying PMC scale setting in the V-scheme, we obtain the PMC scales
\begin{eqnarray}
Q_h =e^{(3/8)}\,m_Q
\label{scaleQCDh}
\end{eqnarray}
for the non-Coulomb correction, and
\begin{eqnarray}
Q_v =2\,v\,m_Q
\label{scaleQCDb}
\end{eqnarray}
for the Coulomb correction. Again, in the V-scheme, $Q_h$ is of order $m_Q$, while $Q_v$ is of order $v\,m_Q$, since the scale $Q_h$ originates from the hard gluon virtual corrections, and $Q_v$ originates from Coulomb rescattering. The physical behavior of the scales does not change using different renormalization schemes. We note that the PMC scales in the usual $\overline{\rm MS}$ scheme are different from the scales in the physically-based V-scheme. This difference is due to the convention used in defining the $\overline{\rm MS}$ scheme. The PMC predictions eliminate the dependence on the renormalization scheme; this is explicitly displayed in the form of ``commensurate scale relations" (CSR)~\cite{Brodsky:1994eh, Lu:1992nt}.

\begin{figure}[htb]
\centering
\includegraphics[width=0.40\textwidth]{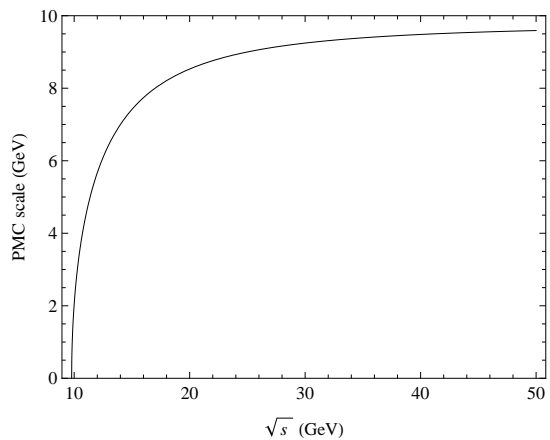}
\caption{The PMC scale $Q_v$ versus the center-of-mass energy $\sqrt{s}$ for the $b$ quark pair production in the V-scheme. $m_Q=4.89$ GeV. }
\label{figPMCscalebeta}
\end{figure}

Taking $m_Q=4.89$ GeV for the $b$ quark pair production, we obtain $Q_h=7.11$ GeV for the non-Coulomb correction, and its value is larger than the conventional choice $\mu_r=m_Q$. For the Coulomb correction, we present its PMC scale $Q_v$ versus the center-of-mass energy $\sqrt{s}$ for the $b$ quark pair production in the V-scheme in Fig.(\ref{figPMCscalebeta}). The exponent disappears in Eq.(\ref{scaleQCDb}) compared to the scale in Eq.(\ref{scaleQCDbms}) in the $\overline{\rm MS}$ scheme. The scale $Q_v$ becomes soft for $v\rightarrow0$, and $Q_v\rightarrow2m_Q$ for $v\rightarrow1$, yielding the correct physical behavior.

\begin{figure}[htb]
\centering
\includegraphics[width=0.40\textwidth]{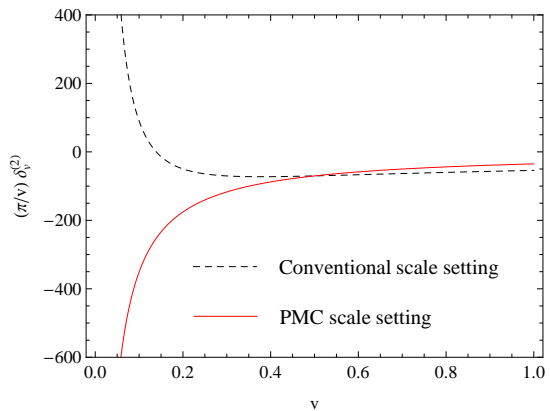}
\caption{The Coulomb terms of the form $(\pi/v)\,\delta^{(2)}_{v}$ in the V-scheme for the $b$ quark pair production, where $\delta^{(2)}_v=(\delta^{(2)}_{v,in}|_V+\delta^{(2)}_{v,n_f}|_V\,n_f)$ is for conventional scale setting and $\delta^{(2)}_v=\delta^{(2)}_{v,\rm sc}|_V$ is for PMC scale setting. }
\label{figCoedetabV2}
\end{figure}

As in the case of the $\overline{\rm MS}$ scheme, the $v$-dependent behavior of the coefficients $\delta^{(2)}_h$ of the non-Coulomb correction in the V-scheme using conventional and PMC scale settings is the same. For the Coulomb correction, the high-order Coulomb term of the form $(\pi/v)^2\,\delta^{(2)}_{v^2}$ are not finite for $v\rightarrow0$ before and after using the PMC. This is because the coefficient $\delta^{(2)}_{v^2}$ of the high-order Coulomb term is independent of the nonconformal $\{\beta_i\}$-terms and the logarithmic term $\ln(v)$. After absorbing the nonconformal $\beta_0$-term using the PMC, the behavior of the Coulomb term of the form $(\pi/v)\,\delta^{(2)}_{v}$ is dramatically changed. More explicitly, the Coulomb terms of the form $(\pi/v)\,\delta^{(2)}_{v}$ in the V-scheme using conventional and PMC scale settings are presented in Fig.(\ref{figCoedetabV2}). When the quark velocity $v\rightarrow0$, the Coulomb term is $(\pi/v)\delta^{(2)}_v\rightarrow+\infty$ due to the presence of the term $-\ln v/v$ using conventional scale setting. After applying PMC scale setting, the logarithmic term $\ln(v)$ vanishes in the coefficient $\delta^{(2)}_v$; the Coulomb term is $(\pi/v)\delta^{(2)}_v\rightarrow-\infty$ due to the term $-(\pi/v)$. Thus, multiplying by the $v$-factor in the LO cross section $\sigma^{(0)}$, the Coulomb term $\sigma^{(0)}(\pi/v)\,\delta^{(2)}_v$ for $v\rightarrow0$ is not finite using conventional scale setting, but it is finite after using PMC scale setting. It is noted that the quark pair and lepton pair productions in $e^+e^-$ annihilation near the threshold region should show similar physical behavior. This dramatically different behavior of the $(\pi/v)\delta^{(2)}_v$ between conventional and PMC scale settings near the threshold region should be checked in QED.

\section{The QED process of the lepton pair production}
\label{sec:3}

Similar to  quark pair production, the lepton pair production cross section for the QED process $e^{+}e^{-}\rightarrow\gamma^{*}\rightarrow l\bar{l}$ is expanded in the QED coupling constant $\alpha$. The cross section can also be divided into the non-Coulomb and Coulomb parts. As in the Eq.(\ref{eqpairNNLO}), the corresponding perturbative coefficients for the lepton pair production cross section are~\cite{Czarnecki:1997vz, Hoang:1995ex, Hoang:1997sj},
\begin{eqnarray}
\sigma^{(0)}=\frac{4}{3}\frac{\pi\,\alpha^2}{s}\,\frac{v\,(3-v^2)}{2},
\end{eqnarray}
\begin{eqnarray}
\delta^{(1)}_h=-4\,C_F,~\delta^{(1)}_{v}=\frac{C_F\,\pi}{2},
\end{eqnarray}
\begin{eqnarray}
\delta^{(2)}_{h,in}&=&{1\over36}\,C_F\,(-16(-11+\pi^2)\,T_R + C_F\,(351+6\pi^4 \nonumber\\
&&-36\zeta_3 + \pi^2\,(-70+48\ln2)) - 24\,C_F\,\pi^2\ln v), \nonumber\\
\delta^{(2)}_{h,n_f}&=&{11\,C_F\,T_R\over9}, \nonumber\\
\delta^{(2)}_{v,in}&=&-2\,C_F^2\,\pi, \nonumber\\
\delta^{(2)}_{v,n_f}&=&{1\over18}\,C_F\,\pi\,T_R\,(-5+6\ln(2v)), \nonumber\\
\delta^{(2)}_{v^2}&=&{C_F^2\,\pi^2\over12}.
\end{eqnarray}
The one-loop correction coefficients $\delta^{(1)}_h$ and $\delta^{(1)}_{v}$ and the two-loop correction coefficients $\delta^{(2)}_{h,n_f}$, $\delta^{(2)}_{v,n_f}$ and $\delta^{(2)}_{v^2}$ have the same form in QCD and QED with only some replacements: $C_A=3$, $C_F=4/3$ and $T_R=1/2$ in QCD and $C_A=0$, $C_F=1$ and $T_R=1$ in QED.

By using the PMC, the vacuum polarization corrections can be absorbed into the QED running coupling:
\begin{eqnarray}
\alpha(Q) = \alpha\left[1 + \left({\alpha\over\pi}\right)\sum^{n_f}\limits_{i=1}{1\over3}\left(\ln\left({Q^2\over m^2_i}\right)-{5\over3}\right)\right],
\end{eqnarray}
where $m_i$ is the mass of the light virtual lepton, and it is far smaller than the final state lepton mass $m_l$. We then obtain
\begin{eqnarray}
\sigma &=& \sigma^{(0)}\left[1+\delta^{(1)}_h\,{\alpha(Q_h)\over\pi} + \delta^{(2)}_{h,\rm in}\,\left({\alpha(Q_h)\over\pi}\right)^2 \right. \nonumber\\
&&\left. + \left(\frac{\pi}{v}\right)\,\delta^{(1)}_v\,{\alpha(Q_v)\over\pi} + \left(\frac{\pi}{v}\right)\,\delta^{(2)}_{v,\rm in}\,\left({\alpha(Q_v)\over\pi}\right)^2\right. \nonumber\\
&&\left. + \left(\frac{\pi}{v}\right)^2\,\delta^{(2)}_{v^2}\,\left({\alpha(Q_v)\over\pi}\right)^2 + {\cal O}(\alpha^3)\right].
\label{QED:afPMC}
\end{eqnarray}
The resulting PMC scales can be written as
\begin{eqnarray}
Q_i = m_l\,\exp\left[{5\over6}+{3\over2}\,{\delta^{(2)}_{i,n_f}\over\delta^{(1)}_i}\right],
\end{eqnarray}
where, $i=h$ and $v$ stand for the non-Coulomb and Coulomb corrections, respectively. Taking $C_A=0$, $C_F=1$ and $T_R=1$ for QED, the PMC scales are
\begin{eqnarray}
Q_h =e^{(3/8)}\,m_l
\label{scaleQEDh}
\end{eqnarray}
for the non-Coulomb correction, and
\begin{eqnarray}
Q_v =2\,v\,m_l
\label{scaleQEDb}
\end{eqnarray}
for the Coulomb correction. Since the scales $Q_h$ stem from the hard virtual photons corrections and $Q_v$ originates from the Coulomb rescattering, $Q_h$ is of order $m_l$ and $Q_v$ is of order $v\,m_l$. The scales show the same physical behavior from QCD to QED after using PMC scale setting. It is noted that the PMC scales in Eqs.(\ref{scaleQCDh}) and (\ref{scaleQCDb}) for QCD in the V-scheme coincide with the scales in Eqs.(\ref{scaleQEDh}) and (\ref{scaleQEDb}) for QED, respectively. This scale self-consistency shows that the PMC method in QCD agrees with the standard Gell-Mann-Low method~\cite{GellMann:1954fq} in QED. The V-scheme provides a natural scheme for the QCD process for the quark pair productions.

In the following, we take the $\tau$ lepton pair production as an example to make a detailed analysis near the threshold region. Taking $m_\tau=1.777$ GeV~\cite{Tanabashi:2018oca}, we obtain the scale $Q_h = 2.59$ GeV, which is larger than $m_\tau$ for the non-Coulomb correction. For the Coulomb correction, as in the case of QCD, the scale becomes soft for $v\rightarrow0$ and $Q_v\rightarrow2m_l$ for $v\rightarrow1$. The PMC scales thus rigorously yield the correct physical behavior for the lepton pair production near the threshold region.

\begin{figure}[htb]
\centering
\includegraphics[width=0.40\textwidth]{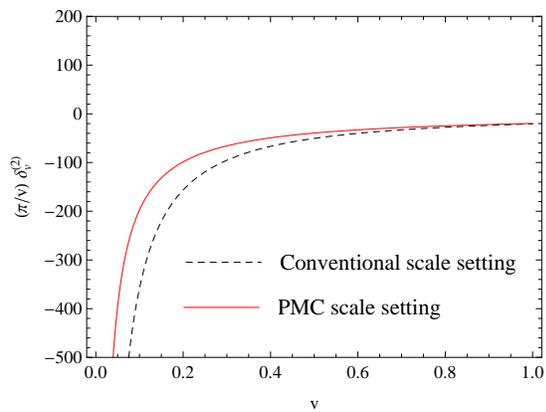}
\caption{The Coulomb terms of the form $(\pi/v)\,\delta^{(2)}_{v}$ for the $\tau$ lepton pair production, where $\delta^{(2)}_v=(\delta^{(2)}_{v,in}+\delta^{(2)}_{v,n_f}\,n_f)$ is for conventional scale setting and $\delta^{(2)}_v=\delta^{(2)}_{v,\rm in}$ is for PMC scale setting. }
\label{figCoeQEDb2}
\end{figure}

For the non-Coulomb correction, the $v$-dependent behavior of the coefficients $\delta^{(2)}_h$ using conventional and PMC scale settings is the same, as in the case of QCD. For the Coulomb correction, the high-order Coulomb term of the form $(\pi/v)^2\,\delta^{(2)}_{v^2}$ are not finite for $v\rightarrow0$ before and after using the PMC. The Coulomb terms of the form $(\pi/v)\,\delta^{(2)}_{v}$ using conventional and PMC scale settings are presented in Fig.(\ref{figCoeQEDb2}). It is noted that in different from the case of QCD, when the quark velocity $v\rightarrow0$, the Coulomb terms are $(\pi/v)\delta^{(2)}_v\rightarrow-\infty$ due to the presence of the term $\ln v/v$ using conventional scale setting, and the term $-1/v$ using PMC scale setting. Multiplying by the $v$-factor in the LO cross section $\sigma^{(0)}$, the Coulomb term for $v\rightarrow0$ is $(\sigma^{(0)}\,(\pi/v)\,\delta^{(2)}_v)\rightarrow-\infty$ using conventional scale setting, and is a finite using PMC scale setting. Thus, we can see from Figs.(\ref{figCoedetab2}), (\ref{figCoedetabV2}) and (\ref{figCoeQEDb2}) that after using the PMC, the behavior of the coefficients $\delta^{(2)}_v$ is dramatically changed; the coefficients $\delta^{(2)}_v$ in the threshold region are not finite using conventional scale setting, and are finite using PMC scale setting for both QCD and QED.

\section{Summary}
\label{sec:4}

Heavy fermion pair production in $e^+e^-$ annihilation is a fundamental process in the SM. However, the conventional procedure of simply setting the renormalization scale as $\mu_r=m_f$ violates the physical behavior of the reaction and leads to the unreliable predictions near the threshold region. In contrast, the PMC scale-setting method provides a self-consistent analysis, and reveals the correct physical behavior of the scale for the heavy fermion pair production near the threshold region, both in  QCD and QED.
\begin{itemize}
\item It is remarkable that two distinctly different scales are determined for the heavy fermion pair production near the threshold region using the PMC. The scale determined for the hard virtual correction is of order the fermion mass $m_f$; the scale determined for the Coulomb rescattering is of order $v\,m_f$, which becomes soft for $v\rightarrow0$. Thus, PMC scale-setting provides a rigorous method for setting unambiguously the renormalization scale as function of the quark velocity $v$, reflecting the virtuality of the propagating gluons (photons) for the QCD (QED) processes.
\item For the non-Coulomb correction of the fermion pair production, the contributions will be suppressed in the threshold region. For the Coulomb correction, the contribution in the threshold region is not finite using conventional scale setting. A resummation of the logarithmically enhanced terms is thus required. After using PMC scale setting, the logarithmic terms $\ln(v)$ vanishes in the coefficient $\delta^{(2)}_v$ from QCD to QED, and thus the coefficient $\delta^{(2)}_v$ is finite in the threshold region.

\item The V-scheme provides a natural scheme for the QCD calculation for the quark pair production. After converting the QCD calculation from the $\overline{\rm MS}$ scheme to the V-scheme, the resulting PMC predictions in the Abelian limit are consistent with the results of QED. The scales are $Q_h =e^{(3/8)}\,m_f$ for the hard virtual correction and $Q_v =2\,v\,m_f$ for the Coulomb rescattering for both QCD and QED. The PMC scales for QCD and QED are identical after applying the relation between PMC scales: $Q^2_{\rm QCD}/Q^2_{\rm QED}=e^{-5/3}$; this factor converts the scale underlying predictions in the $\overline{\rm MS}$ scheme used in QCD to the scale of the V-scheme conventionally used in QED~\cite{Brodsky:1994eh}. We emphasize that the predictions based on the conventional scale-setting method are incorrect when applied to the Abelian theory. The renormalization scale in QED can be set unambiguously by using the Gell-Mann-Low method. The PMC scale-setting method in QCD reduces correctly in the Abelian limit $N_C\rightarrow0$ to the Gell-Mann-Low method. This consistency provides rigorous support for the PMC scale-setting method.
\end{itemize}

\hspace{1cm}

{\bf Acknowledgements}: S. Q. W. thanks the SLAC theory group for kind hospitality. L. D. G. wants to thank the SLAC theory group for its kind hospitality and support. This work was supported in part by the Natural Science Foundation of China under Grants No. 11625520, No. 11705033, and No. 11905056; by the Project of Guizhou Provincial Department under Grant No. KY[2017]067; and by the Department of Energy Contract No.DE-AC02-76SF00515. SLAC-PUB-17512.

\end{document}